\documentstyle[stwol,epsf]{article}

\def\Journal#1#2#3#4{{#1} {\bf #2}, #3 (#4)}
\def\ibid#1#2#3{{\bf #1}, #2 (#3)}

\def\NPB{{\em Nucl.Phys.} B}
\def\PLB{{\em Phys.Lett.}  B}

\def\SJNP{\em Sov.J.Nucl.Phys.}
\def\SJETP{\em Sov.Phys.JETP}

\def\be{\begin{equation}}
\def\ee{\end{equation}}
\def\bea{\begin{eqnarray}}
\def\eea{\end{eqnarray}}

\def \be  {\begin{equation}}
\def \ee  {\end{equation}}
\def \ba  {\begin{eqnarray}}
\def \ea  {\end{eqnarray}}
\def \baa {\begin{eqnarray*}}
\def \eaa {\end{eqnarray*}}
\newcommand\re[1]{(\ref{#1})}

\newcommand{\as}{\ifmmode\alpha_{\rm s}\else{$\alpha_{\rm s}$}\fi}
\def \e {\mbox{e}}
\def \ln {\mbox{ln}}
\def \CO {{\cal O}}
\def \CA {{\cal A}}
\def \CH {{\cal H}}
\newcommand \ci [1] {\cite{#1}}
\def \lab #1 {\label{#1}}
\newcommand \widebar [1] {\overline{#1}}

\def\II{\hbox{{1}\kern-.25em\hbox{l}}}
\font\cmss=cmss10 \font\cmsss=cmss10 at 11pt
\def\inbar{\,\vrule height1.5ex width.4pt depth0pt}
\def\IC{\relax\hbox{$\inbar\kern-.3em{\rm C}$}}
\def\IZ{\relax\ifmmode\mathchoice
{\hbox{\cmss Z\kern-.4em Z}}{\hbox{\cmss Z\kern-.4em Z}}
{\lower.9pt\hbox{\cmsss Z\kern-.4em Z}}
{\lower1.2pt\hbox{\cmsss Z\kern-.4em Z}}\else{\cmss Z\kern-.4em Z}\fi}

\def \Im {\mbox{Im\,}}
\def\IR{{\hbox{{\rm I}\kern-.2em\hbox{\rm R}}}}

\begin{document}
\begin{titlepage}
\def\thefootnote{\fnsymbol{footnote}}
\thispagestyle{empty}
\hfill\parbox{50mm}{{\sc LPTHE--Orsay--96--89} \par
                         hep-ph/9610454      \par
                         October, 1996}
\vspace*{35mm}
\begin{center}
{\LARGE QCD Pomeron as a soliton wave}
\par\vspace*{15mm}\par
{\large G.~P.~Korchemsky}
\footnote{E-mail: korchems@qcd.th.u-psud.fr}
\par\bigskip\par\medskip

{\em Laboratoire de Physique Th\'eorique et Hautes Energies
\footnote{Laboratoire associ\'e au Centre National de la Recherche
Scientifique (URA D063)}
\\
Universit\'e de Paris XI, Centre d'Orsay, b\^at. 211\\
91405 Orsay C\'edex, France
\\[3mm]
and
\\[3mm]
Laboratory of Theoretical Physics
\\
Joint Institute for Nuclear Research
\\
141980 Dubna, Russia}

\end{center}
\vspace*{12mm}

\begin{abstract}
I review a recent progress in understanding of QCD Pomeron and its
relation to exactly solvable models.
\end{abstract}

\vspace*{35mm}
\begin{center}
{\em To appear in the Proceedings of the \\ 
     28th International Conference on High Energy Physics,\\
     Warsaw, Poland, 25-31 July 1996}
\end{center}

\setcounter{footnote}{0}
\renewcommand{\thefootnote}{\arabic{footnote}}

\end{titlepage}

\title{QCD POMERON AS A SOLITON WAVE}

\author{G.P. KORCHEMSKY}

\address{LPTHE, Universit\'e de Paris XI, 91405 Orsay, France and
LTP, JINR, 141980 Dubna, Russia}

\twocolumn[\maketitle\abstracts{
I review a recent progress in understanding of QCD Pomeron and its
relation to exactly solvable models.
}]


{\bf 1.}
Understanding of the mechanism responsible for the rise of the 
structure function of deeply inelastic scattering (DIS) at small $x$ 
still remains a challenge for QCD. The structure function measures the 
distribution density of partons inside the proton carrying the fraction 
$x$ of the proton momentum and having the transverse size $\sim 1/Q$. At
intermediate $x$ and large $Q^2$, the density of partons is small,  
their interaction is weak being proportional to $\as(Q^2)$ and the proton 
can be thought of as a dilute system of quasifree partons whose distribution 
density is governed by the DGLAP evolution equation. The situation is 
changed however at small $x$. The rise of the structure function as $x\to 0$ 
indicates that the density of partons increases and although the interaction 
between partons is still weak at large $Q^2$ we are not allowed to neglect 
multiparton correlations anymore. 

Thus, at small $x$ we enter into a new regime of QCD, well-known since a 
long time as a Regge asymptotic limit, in which one has to deal with the
dynamics of strongly correlated system of partons. It is widely believed 
that in the Regge limit quarks and gluons should form a new collective 
excitations, Reggeons and Pomerons, and QCD has to be replaced by an effective 
Reggeon field theory~\ci{Gr} (with dual models, QCD string etc. among 
potential candidates). It remains unclear however what are the critical 
values of $Q^2$ and small $x$ at which the Regge dynamics will take over 
under the DGLAP evolution and what is the origin of QCD Pomeron in DIS.

In perturbative QCD approach to the Pomeron \ci{BFKL}, for the sake of 
simplicity, we replace the nonperturbative hadronic states by perturbative 
onium states built from two heavy quarks with mass $M$. In this case, the
hadron-hadron scattering amplitude $\CA(s,t)$ (and the structure function
of DIS, $\sim\Im \CA(s=Q^2/x,t=0)/s$, in particular) can be calculated 
in the Regge limit, $s \gg -t,M^2$, as a sum of Feynman diagrams describing 
the multi-gluon exchanges in the $t-$channel. The result of calculation of 
$\CA(s,t)$ in the leading logarithmic approximation (LLA) was interpreted as 
an emergence of the perturbative Reggeon \ci{BFKL}. Although the Reggeon is 
built from an infinite number of ``bare'' gluons it behaves as a point-like
particle with gluon quantum numbers. In perturbative Regge limit,
the hadrons scatter each other by exchanging Reggeons and their
interaction is described by an effective $S-$matrix. The Reggeons propagate 
in the $t-$channel between two hadrons and interacting with each other they 
change their 2-dim transverse momenta, $k_\perp$, but preserve the strong 
ordering of the longitudinal momenta, $k_\pm$. As a result \ci{BFKL}, the 
Reggeon rapidity $y=\ln\frac{k_+}{k_-}$ can be interpreted as a ``time'' in the
$t-$channel and evolution of the system of interacting Reggeons in the
$t-$channel is governed by the effective $(2+1)-$dim Reggeon 
$S-$matrix whose exact expression is unknown yet. In what follows we will 
evaluate the scattering amplitudes in the generalized LLA \ci{BKP}. In this 
approximation one preserves unitarity of the $S-$matrix in the direct 
channels but not in the subchannels.

{\bf 2.}
In the generalized LLA the hadron-hadron scattering amplitude is given
by the sum of effective Reggeon ladder diagrams, in which an arbitrary 
number of Reggeons propagate in the $t-$channel between two hadrons. 
The interaction between Reggeons is elastical and pair-wise. The number 
of Reggeons in the $t-$channel, $N$, is conserved and for given $N$ the 
scattering amplitude satisfies the BKP equation \ci{BKP}. The solutions
of this Bethe-Salpeter like equation define the color-singlet compound 
states built from $N$ interacting Reggeons, perturbative QCD Pomerons
and Odderon. Their contribution to the scattering amplitude takes the 
standard Regge form
$$
\CA(s,t)=is\sum_{N=2}^\infty \as^{N-2} \sum_{\{q\}} 
\beta_{N\to A}^{\{q\}}(t) \beta_{N\to B}^{\{q\}}(t)
s^{E_{N,\{q\}}}
$$
where indices $A$ and $B$ refer to the scattered hadrons. Here, the 
energy of the $N$ Reggeon compound state, $E_{N,\{q\}}$, is defined
as an eigenvalue of the $N$ Reggeon Hamiltonian, $\CH_N$, acting on 
the 2-dim transverse Reggeon momenta
\be
\CH_N |\chi_{N,\{q\}}\rangle = E_{N,\{q\}} |\chi_{N,\{q\}}\rangle
\lab{S}
\ee
with $\{q\}$ being some set of quantum numbers parameterizing all possible
solutions. The residue functions $\beta_{N\to A(B)}^{\{q\}}=\langle A(B)
|\chi_{N,\{q\}}\rangle$ measure the coupling of the $N$ Reggeon state to 
the hadronic states. For given $N$ the scattering amplitude $\CA(s,t)$ 
gets a leading contribution from the Reggeon states with the maximal 
energy
\be
\alpha_N-1={\rm max}_{\{q\}} E_{N,\{q\}} 
\lab{int}
\ee
and $\alpha_N$ can be interpreted as an intercept of a Regge trajectory.
Its character (Regge cut or pole) depends on the distribution density of 
the energy levels $E_{N,\{q\}}$ close to $\alpha_N-1$. 

Being rewritten in the configuration space, the Schr\"odinger equation 
\re{S} describes the dynamics of $N$ pair-wise interacting Reggeons on 
the 2-dim plane of impact parameters $b=(x,y)$ with the effective QCD 
Hamiltonian $\CH_N$ having the following remarkable properties
in the multi-color limit, $N_c\to\infty$ and $\as N_c={\rm fixed}$.
Firstly, the dynamics of Reggeons in holomorphic, $z=x+iy$, and 
antiholomorphic, $\bar z=x-iy$, directions turns out to be independent 
on each other and $\CH_N$ splits into the sum of mutually commuting 
holomorphic and antiholomorphic 1-dim hamiltonians \ci{L}
$$
\CH_N=\frac{\as N_c}{4\pi}\left(H_N + \widebar H_N\right)\,,
$$
where $H_N$ and $\widebar H_N$ describe the nearest-neighbour
interaction between $N$ Reggeons on the line with (anti)holomorphic 
coordinates $z_k$ and $\bar z_k$ ($k=1,...,N$) and periodic boundary
conditions. Secondly, the operator $H_N$ (and $\widebar H_N$) was 
identified as a hamiltonian of completely integrable 1-dim XXX Heisenberg
magnet for a noncompact $SL(2,\IC)$ spin $s=0$ and with the number 
of sites equal to the number of Reggeons \ci{FKL}. As a result, the system
of $N$ Reggeons contains the family of $N-1$ mutually commuting 
holomorphic conserved charges 
$$
q_k=\sum_{N \ge j_1 ... \ge j_k \ge 1} 
i^k z_{j_1j_2} z_{j_2j_3} ... z_{j_kj_1}
\partial_{j_1}\partial_{j_2} ... \partial_{j_N}
$$
with $z_{jk}=z_j-z_k$
and their eigenvalues together with the corresponding antiholomorphic
eigenvalues form the set of quantum numbers of $N$ Reggeon
compound state. Finally, the spectrum of the $N$ Reggeon states is 
defined as
\ba
E_{N,\{q\}}=\frac{\as N_c}{4\pi}\left(\varepsilon_{N,\{q\}}
+\widebar\varepsilon_{N,\{q\}}\right)\,,
\lab{en}
\\
\nonumber
\chi_{N,\{q\}}(z,\bar z)=\varphi_{N,\{q\}}(z)\ 
\widebar\varphi_{N,\{q\}}(\bar z)\,,
\ea
where $\varepsilon_{N,\{q\}}$ and $\varphi_{N,\{q\}}(z)$
are the (holomorphic) energy and the wave function of the 
Heisenberg magnet. The intercept \re{int} can be identified as a 
ground state energy of the XXX Heisenberg magnet with $N$ sites.
 
{\bf 3.}
To find the explicit expression for $E_{N,\{q\}}$ from \re{en} 
one has to derive the quantization conditions for $q_2$, $...$, 
$q_N$ and establish the dependence of $\varepsilon_{N,\{q\}}$ on 
their eigenvalues. This can be done by using the generalized
Bethe Ansatz developed in \ci{K} and based on the separation of
variables~\ci{SoV}. The operators $q_k$ act on holomorphic coordinates
of $N$ Reggeons and their diagonalization is reduced to solving of
a complicated system of $N$ coupled Schr\"odinger equations for
eigenvalues of $q_k$. Instead of dealing with this system we perform 
a unitary transformation, $z_k \to x_k=U^\dagger z_k U$,
in order to replace the original set of Reggeon coordinates $z_k$ by 
a new set of separated variables $x_k$ in terms of which the same 
system of equations decouples into $N$ independent Schr\"odinger 
equations and the Reggeon wave function takes the following
factorized form in new coordinates~\ci{pre}
\be
\varphi_{_{N,\{q\}}}(x_1,..,x_N)=Q(x_1)...Q(x_{N-1})\, \e^{iPx_N}\,,
\lab{wf}
\ee
where $P$ is the total (holomorphic) momentum of the Reggeon state, 
$x_N=\frac1{N}\sum_k z_k$ is the center-of-mass coordinate and the function 
$Q(x)$ satisfies the Baxter equation \ci{K,pre}
\be
x^{-N} \Lambda(x)\, Q(x) = Q(x+i)+Q(x-i)
\lab{L}
\ee
where $N$ is the number of Reggeons inside the compound state, $q_k$ are the
corresponding quantum numbers and $\Lambda(x)=2x^N + q_2 x^{N-2}+...+q_N$.
Having solved the Baxter equation one can obtain the wave function of
the $N$ Reggeon compound state \re{wf} and calculate its holomorphic energy 
using the relation \ci{K}
$$
\varepsilon_{N,\{q\}}=i\frac{d}{dx}\ln\frac{Q(x-i)}{Q(x+i)}\bigg|_{x=0}\,.
$$
This expression determines the dependence of the energy on the quantum 
numbers $q_k$. The Reggeon wave function belongs to the principal series 
representation of the $SL(2,\IC)$ group and the conserved charges $q_k$ 
can be interpreted as higher Casimir operators. In particular, $q_2=-h(h-1)$ 
is the quadratic Casimir.  Its eigenvalue $h$ takes the following quantized 
values
\be
h=\frac{1+m}2+i\nu\,,\qquad m = \IZ\,,\quad \nu=\IR\,,
\lab{h-q}
\ee
which define the conformal weight of the $N$ Reggeon state. Here,
integer $m$ is the Lorentz spin of the Reggeon state $\chi_{N,\{q\}}$, 
corresponding to the rotations in the 2-dimensional impact 
parameter space. The quantization conditions for the remaining charges 
$q_3$, $...$, $q_N$ are much more involved \ci{K}.

To find the solution to the Baxter equation \re{L} one has to specify 
the appropriate boundary conditions on the function $Q(x)$. Their general 
form was not found yet (for recent progress see Ref.~\ci{RW}), except of 
the subclass of polynomial solutions of the Baxter equation \ci{K}, 
corresponding to the special values of quantized conformal weight \re{h-q}, 
$h = \IZ_+$ and $h \ge N$, and leading to the following expression
$$
Q(x)=x^N \prod_{j=1}^{h-N} (x-\lambda_j)\,.
$$
where roots $\{\lambda_j\}$ satisfy the Bethe equations for the
XXX magnet of spin $s=0$. Using polynomial solutions one can
calculate the spectrum of $N$ Reggeon states and then analytically
continue the resulting expressions to arbitrary quantized values of the 
conformal weight. 

{\bf 4.}
The study of the polynomial solutions reveals the following interesting 
properties of the Baxter equation \ci{K}. For given integer 
$h\ge N$, the space of polynomial solutions is finite-dimensional. The 
possible values of roots $\lambda_k$, as well as the values of quantized $q_k$
and the energy $\varepsilon_N$, turn out to be real and simple
$$
\Im \lambda_j = \Im q_k = \Im \varepsilon_N = 0
$$
and they can be parameterized by the set of integers 
$\{n\}=n_1$, $...$, $n_{N-2}$
\be
q_k=q_k(h;\{n\})\,,\qquad
\varepsilon_N=\varepsilon_N(h;\{n\})
\lab{n}
\ee
such that $n_1,..., n_{N-2} \ge 0$ and $\sum_{k=1}^{N-2} n_k \le h-N$.
As an example, the quantized values of $q_3$ and holomorphic energy 
$\varepsilon_3$ for $N=3$ Reggeon compound states obtained from numerical 
solutions of the Baxter equation for integer $h$ are shown by dots in Figs.1 
and 2, respectively.%
\begin{figure}[ht]
\vspace*{-13mm}
\centerline{\epsfysize=8cm\epsfxsize=9.5cm\epsffile{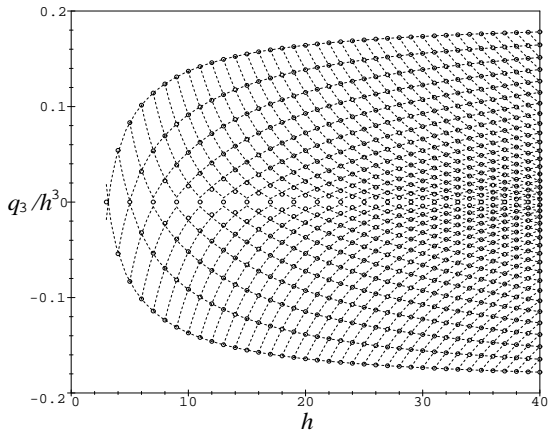}}
\vspace*{-14mm}
\caption{Quantized $q_3$ for the $N=3$ Reggeon states}
\end{figure}
\begin{figure}[ht]
\vspace*{-17mm}
\centerline{\epsfysize=8cm\epsfxsize=9.5cm\epsffile{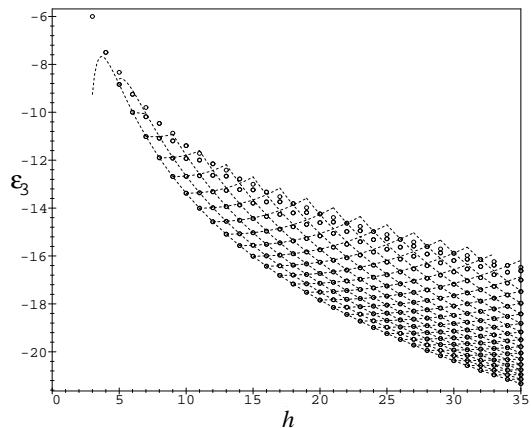}}
\vspace*{-14mm}
\caption{Quantized energy of the $N=3$ Reggeon states}
\end{figure}
The Baxter equation has the form of a discretized 1-dim Schr\"odinger equation
and one can apply the WKB expansion \ci{pre} to find its asymptotic solution as
$$
Q(x)=\exp(iS_{_{\rm WKB}}(x))\,,\quad
S_{_{\rm WKB}}=S_0+ S_1+ ...\,.
$$
The leading term $S_0(x)$ defines the semiclassical Reggeon dynamics in the 
collective coordinates $x_k$. To describe the classical trajectories of $N$
Reggeons it is convenient to introduce the complex curve $w$
$$
\Gamma_N:\quad
\omega+\frac1{\omega}=x^{-N} \Lambda(x)
$$
with $\Lambda(x)$ defined in \re{L}, which is a single-valued function
of complex $x$ on the hyperelliptic Riemann surface obtained by 
gluing together two sheets of the complex $x-$plane along the cuts running
between the branch points $\sigma_j$ defined as
$$
\sigma_j^{-N} \Lambda(\sigma_j) = \pm 2\,.
$$
The genus of the Riemann surface $\Gamma_N$, $g=N-2$, depends on the
number of Reggeons inside the compound state. 
Then, the Reggeon momentum in the separated coordinates is given by 
$p=\ln |\omega(x)|$ and the action $S_0$ can be obtained as an integral
of a meromorphic differential on $\Gamma_N$
$$
S_0=\int^Q dx \,\ln\, \omega\cong - \int^Q x\,\frac{d\omega}{\omega}
$$
with $Q=(x,\pm)$ being the point on the Riemann surface belonging to either
upper or lower sheet of $\Gamma_N$. In the center-of-mass frame, $x_N=0$,
the classical motion of Reggeons corresponds to the points on $\Gamma_N$ with 
real coordinates and momenta, $(x,p=\ln|\omega|)$. These points belongs to 
the $N-1$ cycles $\alpha_j$ on $\Gamma_N$ which surround the cuts running 
between the branch points $[\sigma_{2j-1},\sigma_{2j}]$ and forming $N-1$ 
compact intervals on the real axis. 
The brach points become the turning points of the classical 
trajectories.

The phase space of $N$ Reggeons is given by the direct product
of the cycles $\alpha_j$ on the Riemann surface $\Gamma_N$ times the 
center-of-mass motion. The set of points $Q_1, ... , Q_{N-1}$ situated 
one each on the $\alpha-$cycles corresponds to the real values of the Reggeon
coordinates $(x_j,p_j)$ and provides the coordinates on the level surface 
$q_k={\rm const}$. 
The conserved charges $q_k$ of the $N$ Reggeon state play the role of 
hamiltonians generating the hamiltonian 
flows of Reggeons on $\Gamma_N$ in ``times'' $\tau_k$. The corresponding 
evolution equations for the Reggeon coordinates have the form \ci{pre}
$$
d\tau_k =\sum_{j=1}^{N-1} \frac{d x_j x_j^{N-k}}
                               {\sqrt{\Lambda^2(x_j)-4x_j^{2N}}}.
$$
They are similar to the soliton equations of the KP/Toda hierarchy and
their solution defines the Reggeon soliton wave in 2-dim plane of the
impact parameters $(z,\bar z)$~\ci{pre}. 
The quantum numbers $q_k$ enter as parameters into the $N$ Reggeon soliton 
waves and their possible values are constrained by the Bohr-Sommerfeld 
quantization conditions
$$
\oint_{\alpha_k} dS_{_{\rm WKB}} = 2\pi n_k
$$
with integers $n_k$ defined in \re{n}. Their solutions define the $N-2$ 
parametric families of curves \re{n}, which can be interpreted \ci{pre}
as corresponding to the Whitham deformation of the Reggeon soliton waves 
in a ``slow'' time $h$, the conformal weight. For large $h$ one can develop 
the asymptotic expansion of $q_k$ and $\varepsilon_N$ in inverse powers of 
the conformal weight~\ci{K}
\ba
q_k &=& h^k \sum_{l=0}^\infty q_k^{_{(l)}}(\{n\}) \, h^{-l}\,,
\nonumber
\\[-2mm]
\varepsilon_N &=& -2N\ln h + \sum_{l=0}^\infty
\varepsilon_N^{_{(l)}}(\{n\}) \, h^{-l}\,,
\lab{as}
\ea
where $k=3$, $...$, $N$ and the coefficients $q_k^{_{(l)}}$ and
$\varepsilon_N^{_{(l)}}$ depend on the integers $n_k$. For $N=2$ Reggeon
state, the BFKL Pomeron, all coefficients are known exactly. For $N=3$ 
Reggeon states $q_3$ and $\varepsilon_3$ were calculated up to $\CO(h^{-8})$ 
order \ci{K} and the results are shown by dotted lines in Figs.1 and 2.  
The asymptotic approximation to the intercept of the $N=3$ Reggeon state,
perturbative Odderon, can be obtained from \re{as} as \ci{K}
$$
\alpha_3^{\rm app} = 1+\frac{\as N_c}{\pi} 2.4131
$$
and it is smaller than the intercept of the BFKL Pomeron \ci{BFKL}, 
$\alpha_2=1+\frac{\as N_c}{\pi}4\ln 2$.

\end{document}